\documentclass[amsthm]{elsart}
\usepackage{yjsco}
\usepackage{natbib}
\usepackage{amsfonts}
\usepackage{amsmath}
\usepackage{amssymb}

\begin{document}

\begin{frontmatter}

\title{Solving Polynomial Equations with Equation Constraints: the Zero-dimensional Case}

\author{Ye Liang}
\address{Max-Plank-Institut f\"ur Informatik, Saarbr\"ucken, Germany}

\ead{wolf39150422@gmail.com}

\begin{abstract}
A zero-dimensional polynomial ideal may have a lot of complex zeros. But sometimes, only some of them are needed. In this paper, for a zero-dimensional ideal $I$, we study its complex zeros that locate in another variety $\textbf{V}(J)$ where $J$ is an arbitrary ideal.

The main problem is that for a point in $\textbf{V}(I) \cap \textbf{V}(J)=\textbf{V}(I+J)$, its multiplicities w.r.t. $I$ and $I+J$ may be different. Therefore, we cannot get the multiplicity of this point w.r.t. $I$ by studying $I + J$. A straightforward way is that first compute the points of $\textbf{V}(I + J)$, then study their multiplicities w.r.t. $I$. But the former step is difficult to realize exactly.

In this paper, we propose a natural geometric explanation of the localization of a polynomial ring corresponding to a semigroup order. Then, based on this view, using the standard basis method and the border basis method, we introduce a way to compute the complex zeros of $I$ in $\textbf{V}(J)$ with their multiplicities w.r.t. $I$. As an application, we compute the sum of Milnor numbers of the singular points on a polynomial hypersurface and work out all the singular points on the hypersurface with their Milnor numbers.

\end{abstract}

\begin{keyword}
 Semigroup order \sep multiplicity \sep zero-dimensional \sep standard basis \sep border basis \sep polynomial equations \sep Milnor number \sep sum
\end{keyword}

\end{frontmatter}

\section{Introduction} \label{introduction}

A crucial difference between linear equations and polynomial equations is that the latter may have many isolated complex solutions. In fact, when the degrees of polynomials and the number of variables become larger, the number of complex solutions of polynomial equations can increase dramatically. But sometimes not all these solutions are of interest. We may only want to get the information of a part of them.

Given two polynomial ideals $I$ and $J$ in $\mathbb{C}[x_1,\ldots,x_n]$ with $I$ zero-dimensional, we want to compute the points in $\textbf{V}(I) \cap \textbf{V}(J)=\textbf{V}(I+J)$ with the multiplicities w.r.t. $I$. Note that a point in $\textbf{V}(I) \cap \textbf{V}(J)$ may have different multiplicities when it is considered as a zero of $I$ or $I+J$. For example, let $I=\langle x^3 \rangle$ and $J=\langle x \rangle$ in $\mathbb{C}[x]$. We can easily obtain that the point $0 \in \mathbb{C}$ has multiplicities $3$ and $1$ w.r.t. $I$ and $I+J=J$, respectively.
Thus, we cannot get the multiplicity of $0$ w.r.t. $I$ by only studying $I+J$. One possible way to solve this problem is that we first compute the common points of $\textbf{V}(I)$ and $\textbf{V}(J)$, then localize $I$ and $\mathbb{C}[x_1,\ldots,x_n]$ at each point of $\textbf{V}(I+J)$ and compute their multiplicities w.r.t. $I$. However, in general, it is not easy to compute the zeros of $I+J$ exactly. Usually, we need a numerical solver and should use floating-point computation. As a result, we cannot get an exact result.

In this paper, the problem is dealt with in another way. We first provide a natural geometric explanation of the localization of $\mathbb{C}[x_1,\ldots,x_n]$ w.r.t. a semigroup order $>$ (cf. Theorem \ref{decomposition} and Corollary \ref{CorollaryMultiplicity}), i.e., for each $1 \leq i \leq n$, $x_i<1$ implies that the complex zeros of $I$ with the $i$-th coordinates nonzero are discarded from $\textbf{V}(I^{ec}) \subset \mathbb{C}^n$. In other words, only the complex zeros $\xi$ of $I$ with $\xi_i =0$ for all $i$ such that $x_i<1$ are kept in $\textbf{V}(I^{ec})$. Based on this geometric view, by renaming constraint polynomials $g_1,\ldots,g_u$ with $\langle g_1,\ldots,g_u \rangle =J$ to new variables $x_{n+1},\ldots,x_{n+u}$, we transform the computation of the zeros of $I$ in $\textbf{V}(J)$ to the computation in a larger polynomial ring $\mathbb{C}[x_1,\ldots,x_{n+u}]$ where Theorem \ref{decomposition} can be used. Now, we need to compute the complex zeros of $I'^{ec}:=\langle I,x_{n+1}-g_1,\ldots,x_{n+u}-g_u \rangle ^{ec}$ w.r.t. a semigroup order $>$ with $x_1>1,\ldots,x_n>1$ and $x_{n+1}<1,\ldots,x_{n+u}<1$.  By using the standard basis method and the Mora normal form algorithm in $\textup{Loc}_{>}(\mathbb{C}[x_1,\ldots,x_{n+u}])$, we can compute the reduced normal form of the monomials in the border $\partial W$ of the order ideal $W$ consisting of the standard monomials. Then, $G:=\{t-\textup{redNF}(t):t \in \partial W \}$ forms a border basis of $I'^{ec}$ in $\mathbb{C}[x_1,\ldots,x_{n+u}]$. From $G$, we can construct multiplication matrices and apply the Chow form method, the rational univariate representation (RUR) method or other methods to work out the complex zeros of $I$ that locate in $\textbf{V}(J)$ with their multiplicities w.r.t $I$. To illustrate this process, we compute the singular points with their Milnor numbers on a polynomial hypersurface in Section \ref{Sec:Computing}.

As far as we know, there is no previous work for solving polynomial equations with equation constraints (no parameters). But there are some related work to our studies of semigroup orders. \cite{Buchberger65} used global orders in setting up his Gr\"obner basis theory. By computing a Gr\"obner basis of $I$, one can calculate all the complex zeros of the ideal. After that, \cite{Mora82} provided an algorithm to compute standard bases by using local orders. A important usage of such a standard basis is to compute the local multiplicity of a given complex zero of $I$. Then, \cite{Robbiano1985} proved that every semigroup order can be described by a suitable matrix. Later, \cite{Greuel96} and \cite{Graebe94} found Mora's algorithm works for any semigroup order.

The rest of this paper is structured as follows. Section \ref{Sec:Pre} is devoted to introducing necessary notions and theorems. In Section \ref{Sec:Classify}, we present a natural geometric explanation of the localization of polynomial rings corresponding to an arbitrary semigroup ordering. In Section \ref{Sec:Partial}, we study partial zeros of $I$ with their multiplicities w.r.t. $I$. Then, in Section \ref{Sec:Computing} we compute these partial solutions. An application in computing Milnor numbers on a polynomial hypersurface is provided at this section as an example. Finally, we make a conclusion in Section \ref{Sec:Conclusion}.

\newtheorem{theorem}{Theorem}[section]
\newtheorem{proposition}[theorem]{Proposition}
\newtheorem{lemma}[theorem]{Lemma}
\newtheorem{definition}[theorem]{Definition}
\newtheorem{observation}[theorem]{Observation}
\newtheorem{corollary}[theorem]{Corollary}
\newtheorem{problem}[theorem]{Problem}
\newtheorem{remark}[theorem]{Remark}
\newtheorem{example}[theorem]{Example}
\newtheorem{assumption}[theorem]{Assumption}

\section{Preliminaries}\label{Sec:Pre}
This section consists of background concepts and theorems. Let $A=\mathbb{C}[x_1,\ldots,x_n]$ and $\textup{T}^{\{x_1,\ldots,x_n\}}=\{x^{\alpha}:\alpha \in \mathbb{Z}^n_{\geq 0} \}$.  The following five definitions come from \cite{Cox05}.

\begin{definition}[Multiplicities]\label{multiplicity}
  Let $I$ be a zero-dimensional ideal in $A$, so that $\textbf{V}(I)$ consists of finitely many points in $\mathbb{C}^n$, and assume that $p=(a_1,\ldots,a_n)$ is one of them. Then the multiplicity of $p$ as a point in $\textbf{V}(I)$ is $\dim_{\mathbb{C}}A_M/IA_M$ where $M={\langle x_1-a_1,\ldots,x_n-a_n \rangle}$.
\end{definition}

\begin{definition}[Semigroup Orders]
An order $>$ on $\mathbb{Z}^n_{\geq 0}$, or equivalently, on $\textup{T}^{\{x_1,\ldots,x_n\}}$ in $A$ is said to be a semigroup order if it satisfies:
\begin{enumerate}
  \item $>$ is a total ordering on $\mathbb{Z}^n_{\geq 0}$;
  \item $>$ is compatible with multiplication of monomials.
\end{enumerate}
\end{definition}

\begin{definition}[Localizations of Rings]
Let $>$ be a semigroup order in $A$ and let $S=\{1+g \in A:g=0 \ \textup{or} \ \textsc{lt}(g)<1\}$. The localization of $A$ w.r.t. $>$ is the ring $$\textup{Loc}_{>}(A)=S^{-1}A=\{f/(1+g): f \in A, \ 1+g \in S\}.$$
\end{definition}

\begin{definition}[Standard Bases]
Let $>$ be a semigroup order and let $I \subset \textup{Loc}_{>}(A)$ be an ideal. A standard basis of $I$ w.r.t. $>$ is a set $\{g_1,\ldots,g_t\} \subset I$ such that $\langle \textsc{lt}(I) \rangle = \langle \textsc{lt}(g_1),\ldots,\textsc{lt}(g_t) \rangle$.
\end{definition}

\begin{definition}[Standard Monomials]\label{Def:StandardMonomial}
Given a semigroup order $>$ and an ideal $I$ in the localization $\textup{Loc}_{>}(A)$ of the ring $A$, we say that a monomial $x^{\alpha}$ is standard if $x^{\alpha} \not \in \langle \textsc{lt}(I) \rangle$.
\end{definition}
The above definition is a generalization of the one in \cite{Cox05}, where the order $>$ is a local order.

Let $f: A \rightarrow S^{-1}A, \ a \mapsto a/1$ be a ring homomorphism where $S$ is a multiplicatively closed subset of $A$. For an ideal $I \subset A$, its extension $I^e$ in $S^{-1}A$ is $I(S^{-1}A)=S^{-1}I$. For an ideal $J \subset S^{-1}A$, its contraction $J^c$ in $A$ is $f^{-1}(J)$. The following two theorems come from \cite{Atiyah1969}.
\begin{theorem}\label{Atiyah}
Let $S$ be a multiplicatively closed subset of $A$, and let $I$ be an ideal. Let $I=\cap^k_{i=1}Q_i$ be a minimal primary decomposition of $I$. Let $P_i$ be the radical of $Q_i$ and suppose the $Q_i$ numbered so that $S$ meets $P_{m+1},\ldots,P_k$ but not $P_1,\ldots,P_m$. Then $$ S^{-1}I=\cap^m_{i=1}S^{-1}Q_i, \quad I^{ec}=(S^{-1}I)^c=\cap^m_{i=1}Q_i$$ and these are minimal primary decompositions.
\end{theorem}

\begin{theorem}\label{Th:index}
Let $A$ be a Noetherian ring, $P$ a maximal ideal of $A$, $Q$ any ideal of $A$. Then the following are equivalent:
\begin{enumerate}
  \item $Q$ is $P$-primary;
  \item $\sqrt{Q}=P$;
  \item $P^k \subset Q \subset P$ for some $k>0$.
\end{enumerate}
\end{theorem}

The contents of border basis come from \cite{KKR}.

\begin{definition}[Order Ideals]\label{Def:OrderIdeal}
A non-empty, finite set of terms $W \subset \textup{T}^{\{x_1,\ldots,x_n\}}$ is called an order ideal if it is closed under forming divisors, i.e., if $t \in W$ and $t'|t$ imply $t' \in W$.
\end{definition}

\begin{definition}[Borders]\label{Def:Border}
Let $W \subset \textup{T}^{\{x_1,\ldots,x_n\}}$ be an order ideal. The border of $W$ is the set $$\partial W=(x_1W \cup \cdots \cup x_nW) \setminus W.$$
\end{definition}

\begin{definition}[$W$-border Prebases]\label{Def:Prebasis}
Let $W=\{t_1,\ldots,t_{\mu}\}$ be an order ideal, and let $\partial W=\{b_1,\ldots,b_{\nu}\}$ be its border. A set of polynomials $G=\{g_1,\ldots,g_{\nu}\}$ is called a $W$-border prebasis if the polynomials have the form $g_j=b_j-\sum_{i=1}^{\mu}\alpha_{ij}t_i$ such that $\alpha_{ij} \in \mathbb{C}$ for $1 \leq i \leq \mu$ and $1 \leq j \leq \nu$.
\end{definition}

\begin{definition}[$W$-border Bases]\label{Def:BorderBasis}
Let $I$ be a zero-dimensional ideal in $A$, $W=\{t_1,\ldots,t_{\mu} \}$ be an order ideal and $G=\{g_1,\ldots, g_{\nu}\}$ be a $W$-border prebasis consisting of polynomials in $I$. We say that the set $G$ is a $W$-border basis of $I$ if the residue classes of $t_1,\ldots,t_{\mu}$ form a $\mathbb{C}$-vector space basis of $A/I$.
\end{definition}

\begin{proposition}\label{Pro:BorderGenerators}
Let $W=\{t_1,\ldots,t_{\mu}\}$ be an order ideal, and let $G$ be a $W$-border basis of $I$. Then $I$ is generated by $G$.
\end{proposition}

The rest of this section comes from \cite{mou07}.

\begin{definition}[Multiplication Matrices]\label{Def:MultiplicationMatrix}
For a polynomial ideal $I \subset A$ and any element $a \in A/I$, we define the map $M_a: A/I \rightarrow A/I$, $b \mapsto ab$. Given a vector space basis of $A/I$, we call the matrix associated to this operator the multiplication matrix w.r.t. the basis and still denote it by $M_a$.
\end{definition}

\begin{definition}[Chow Forms]\label{Def:ChowForm}
For a zero-dimensional polynomial ideal $I \subset A$, the Chow form of $A/I$ is the homogeneous polynomial in $u=(u_0,\ldots,u_n)$ defined by $$C_I(u)=\det(u_0+u_1M_{x_1}+\cdots+u_nM_{x_n}).$$
\end{definition}

\begin{theorem}\label{Th:ChowForm}
The Chow form of $A/I$ is $$C_I(u)=\prod_{\zeta \in \textbf{V}(I)}(u_0+u_1\zeta_1+\cdots+u_n\zeta_n)^{m_{\zeta}},$$ where $\textbf{V}(I)$ is the finite complex zero set of $I$, $\zeta_1, \cdots, \zeta_n$ are the components of $\zeta$, and $m_{\zeta}$ is the multiplicity of $\zeta$.
\end{theorem}

\section{Classification of Semigroup Orders} \label{Sec:Classify}

In this section, we provide a classification of semigroup orders based on their effects on localizing polynomial rings. The following theorem is in the center of all our results. It gives a natural geometric explanation to the localization of $A$ corresponding to a semigroup order $>$, i.e., for a zero-dimensional ideal $I \subset A$, $x_i <1$ implies that the points $\xi \in \textbf{V}(I)$ with $\xi_i \neq 0$ are deleted from $\textbf{V}(I^{ec}) \subset \mathbb{C}^n$. After introducing this theorem we present two corollaries and a definition of the equivalence of semigroup orders.

\begin{theorem}\label{decomposition}
Let $>$ be a semigroup order in $A$ with $x_{j_1}<1,\ldots, x_{j_k} <1$ and $x_{j_{k+1}}>1,\ldots,x_{j_n} >1$ where $(j_1,\ldots,j_n)$ is a permutation of $(1,\ldots,n)$. Let $S=\{1+g:g=0 \vee \textsc{lt}(g)<1,g \in A\}$. Let $I \subset A$ be a zero-dimensional polynomial ideal and $I=\cap^k_{i=1}Q_i$ be its minimal primary decomposition. Let $P_i= \langle x_1-a_{i1},\ldots,x_n-a_{in} \rangle$ be the radical of $Q_i$ and suppose the $Q_i$ numbered so that $a_{ij_1}=a_{ij_2}=\cdots=a_{ij_k}=0$ for and only for the first $m$ $Q_i$.  Then, $S^{-1}I=\cap^m_{i=1}S^{-1}Q_i$ and $I^{ec}=(S^{-1}I)^c=\cap^m_{i=1}Q_i$ are minimal primary decompositions.
\end{theorem}
\begin{pf}
According to Theorem \ref{Atiyah}, we only need to verify that $P_i$ does not meet $S$ if and only if $a_{ij_1}=a_{ij_2}=\cdots=a_{ij_k}=0$.

``$\Leftarrow$" Since $>$ is a semigroup order and $\textsc{lt}(g)<1$ for $1+g\in S \setminus \{1\}$, there exists a $j \in \{j_1,\ldots, j_k\}$ such that $x_j|\textsc{lt}(g)$. Then all the monomials of $g$ can be divided by an element of $\{x_{j_1},\ldots, x_{j_k}\}$. Hence, when substituting $a_{ij_1}=a_{ij_2}=\cdots=a_{ij_k}=0$ to $1+g$ we obtain the result $1$. It means that none of the elements of $S$ vanish under the substitution.

``$\Rightarrow$" Consider $1+\alpha x_j$ with $j \in \{j_1,\ldots, j_k\}$. It locates in $S$ for any $\alpha \in \mathbb{C}$. Since $P_i$ does not meet $S$, we have $a_{ij}=0$.
\end{pf}

\begin{corollary}\label{CorollaryMultiplicity}
In theorem \ref{decomposition}, $I^{ec}=(S^{-1}I)^c$ has the same multiplicity with $I$ at each $P_i$ ($1\leq i \leq m$).
\end{corollary}
\begin{pf}
We know $I^{ec}=(S^{-1}I)^c=\cap^m_{i=1}Q_i$ by Theorem \ref{decomposition}. By Definition \ref{multiplicity}, we only need to check if the localizations of $I^{ec}$ and $I$ are equal at $P_i$ ($1\leq i \leq m$). This is obviously correct according to Theorem \ref{Atiyah}.
\end{pf}

From the view of computing isolated complex roots, semigroup orders can be classified only according to the orders between variables and $1$.

\begin{corollary}\label{equivalence}
For two semigroup orders $>_1$ and $>_2$ in $A$, they have the same effect on the localization of any zero-dimensional ideal $I$, i.e., $(S_1^{-1}I)^c=(S_2^{-1}I)^c$, if and only if for every $1 \leq j \leq n$ we have $x_j>_1 1$ if and only if $x_j>_2 1$.
\end{corollary}

\begin{pf}
``$\Leftarrow$" If for every $1 \leq j \leq n$ we have $x_j>_1 1$ if and only if $x_j>_2 1$, then we have $(S_1^{-1}I)^c=(S_2^{-1}I)^c$ according to Theorem \ref{decomposition}.

``$\Rightarrow$" Suppose there exists a $1 \leq j \leq n$ such that $x_j>_1 1$ (or $x_j<_1 1$) but $x_j<_2 1$ (or $x_j>_2 1$). Then, for the maximal ideal $I=Q=P=\langle x_1-a_1,\ldots,x_n-a_n \rangle$ where $a_j \neq 0$ and $a_i=0$ for $i \neq j$, we have $(S_1^{-1}I)^c=I$ (or $A$) and $(S_2^{-1}I)^c=A$ (or $I$) by Theorem \ref{decomposition}. This is a contradiction.
\end{pf}

Corollary \ref{equivalence} indicates that there exists an equivalence relation among semigroup orders. We make this relation explicit below.
\begin{definition}[Equivalence Relations]
Given two semigroup orders $>_1$ and $>_2$ on $\mathbb{Z}^n_{\geq 0}$, we say that $>_1$ is equivalent to $>_2$ (denoted by $>_1 \sim >_2$) if for every $1 \leq j \leq n$ we have $x_j>_1 1$ if and only if $x_j>_2 1$.
\end{definition}

Corollary \ref{equivalence} says that semigroup orders in the same equivalence class have the same effect in localizing rings. Thus, we only need to choose any representative of the orders in the same equivalence class to compute a standard basis (cf. Example \ref{Example}) for solving polynomial equations. It is easy to see that there are $2^n$ different equivalence classes of semigroup orders on $\mathbb{Z}^n_{\geq 0}$.

The view of Theorem \ref{decomposition} can be very helpful for us to work out the complex zeros of a zero-dimensional ideal $I$ in another variety $\textbf{V}(J)$. We introduce this idea below.

\section{Partial Zeros of Zero-dimensional Ideals} \label{Sec:Partial}

By Theorem \ref{decomposition}, we can only compute the complex zeros of $I$ in $\textbf{V}(\langle x_{j1},\ldots,x_{jk}\rangle)$. For a general case, we want to transform it to this special case and solve it. A natural idea is to rename the generators $g_1,\ldots,g_u$ of the ideal $J$ to new variables $x_{n+1},\ldots,x_{n+u}$. (Similar process has been provided in \cite{Mora95} for different purpose. we will introduce it for our purpose here.) Suppose $I=\langle f_1,\ldots,f_v \rangle$ where $v \geq n$ since $I$ is zero-dimensional. Consider the ideal $\langle f_1,\ldots,f_v,g_1-x_{n+1},\ldots,g_u-x_{n+u} \rangle \subset \mathbb{C}[x_1,\ldots,x_{n+u}]$. We can easily imagine that the complex zeros of this ideal are in one-to-one correspondence with the complex zeros of $I$, and corresponding zeros of the two ideals share the same multiplicities. A proof of a similar result can be found in \cite{FL11Extended}. For completeness, we write down the detail proof of this lemma.

\begin{lemma}\label{CorrespondingZeros}
Specify a zero-dimensional polynomial ideal $I=\langle f_1,\ldots,f_v\rangle \subset A$ and another polynomial ideal $J=\langle g_1,\ldots,g_u \rangle \subset A$. Consider the ideal $I'=\langle f_1,\ldots,f_v,g_1-x_{n+1},\ldots,g_u-x_{n+u} \rangle \subset \mathbb{C}[x_1,\ldots,x_{n+u}]$. Then, we have that the complex zeros of $I'$ and $I$ are in a one-to-one correspondence. The two zeros in each pair have the same first $n$ projections and have the same multiplicity.
\end{lemma}
\begin{pf*}{Proof}
For each complex zero of $I$ we can easily obtain a corresponding complex zero for $I'$ by substituting the zero to $x_{n+1}=g_1, \ldots, x_{n+u}=g_u$. Moreover, different zeros of $I$ yield different zeros of $I'$. On the other hand, for each zero of $I'$, we take the first $n$ components to form a point $p$ in $\mathbb{C}^n$. Then $p$ is obviously a zero of $I$. Hence there exists a one-to-one correspondence between the complex zeros of $I$ and $I'$, and corresponding zeros have the same first $n$ projections. We only need to prove that the two zeros in each pair have the same multiplicity. It is proved by definition as follows.

Let $\mathcal{O}$ and $\mathcal{O}'$ be the rings of rational functions defined at complex zeros $p=(a_1,\ldots,a_n)$ and $p'=(a_1,\ldots, a_n,g_1(p),\ldots,g_u(p))$ of $I$ and $I'$, respectively, i.e., $\mathcal{O}=\{h/g: g,h \in A, g(p) \neq 0 \}$ and $\mathcal{O}'=\{h'/g': g',h' \in \mathbb{C}[x_1,\ldots,x_{n+u}], g'(p') \neq 0 \}$. Let $I\mathcal{O}$ be the ideal generated by $I$ in $\mathcal{O}$. Let $\varphi : \mathcal{O}'\longrightarrow \mathcal{O}/I\mathcal{O}$ be a homomorphism such that $$\varphi(h'/g')=\frac{h'(x_1,\ldots, x_n, g_1,\ldots,g_u)}{g'(x_1,\ldots, x_n, g_1,\ldots,g_u)}+I\mathcal{O}$$ where $h',g'\in \mathbb{\mathbb{C}}[x_1,\ldots,x_{n+u}]$ and $g'(p') \neq 0$. It is easy to see this homomorphism is surjective. For any $q'\in \mathcal{O}'$, if
$\varphi(q')=q'(x_1,\ldots, x_n, g_1,\ldots,g_u)+I\mathcal{O}=I\mathcal{O}$, then
$$q'(x_1,\ldots, x_n, g_1,\ldots,g_u)=\frac{h'(x_1,\ldots, x_n, g_1,\ldots,g_u)}{g'(x_1,\ldots, x_n, g_1,\ldots,g_u)}\in I\mathcal{O}.$$ Hence $h'(x_1,\ldots, x_n, g_1,\ldots,g_u)\in
I\mathcal{O} \subset I'\mathcal{O}'$. Each term in
$\textup{T}^{\{x_1,\ldots,x_{n+u}\}}$ can be written as
\begin{eqnarray*}
\qquad \quad \quad \  &&x_{1}^{u_1} \cdots x_{n}^{u_n} x_{n+1}^{v_1} \cdots x_{n+u}^{v_u}\\
& = & x_{1}^{u_1} \cdots x_{n}^{u_n} (g_{1}+(x_{n+1}-g_{1}))^{v_1} \cdots (g_{u}+(x_{n+u}-g_{u}))^{v_u}\\
& = & x_{1}^{u_1} \cdots x_{n}^{u_n} g_{1}^{v_1} \cdots
g_{u}^{v_u}+\sum_{j=1}^u r_j(x_{n+j}-g_j)
\end{eqnarray*}
where $r_j \in \mathbb{C}[x_1,\ldots,x_{n+u}]$. Hence,
\begin{equation*}
h'(x_1,\ldots,x_{n+u})=h'(x_1,\ldots,x_n,g_1,\ldots,g_u)+\sum_{j=1}^u r_j^*(x_{n+j}-g_j)\in I'\mathcal{O}'
\end{equation*}
where $r_j^* \in \mathbb{C}[x_1,\ldots,x_{n+u}]$. Therefore, $q'=h'/g' \in I'
\mathcal{O}'$ and $\ker(\varphi)\subset I' \mathcal{O}'$. Conversely, for every $q' \in I'\mathcal{O}'$ there exists a $g'\in
\mathbb{C}[x_1,\ldots,x_{n+u}]$ with $g'(p') \neq 0$ such that $g'q'
\in I'$. Hence $g'(p')\neq 0$ and
 $(g'q')(x_1,\ldots,x_n,g_1,\ldots,g_u) \in I$. Therefore, $\varphi (q') =I \mathcal{O} $ and $\ker(\varphi)\supset I' \mathcal{O}'$.
 We have $\ker(\varphi)= I' \mathcal{O}'$,
 and $\mathcal{O}'/I' \mathcal{O}' \cong \mathcal{O}/I \mathcal{O}$. This ring isomorphism is also an isomorphism of vector spaces over $\mathbb{C}$.
 Consequently, $\textup{dim}(\mathcal{O}'/I' \mathcal{O}') =\textup{dim}( \mathcal{O}/I
 \mathcal{O})$, i.e., $p$ and $p'$ have the same multiplicity.
\end{pf*}

In the following theorem, we transform the problem of computing the complex zeros of $I$ that locate in $\textbf{V}(J)$ with multiplicities w.r.t. $I$ to a problem in a larger ring that Theorem \ref{decomposition} can be used to solve.

\begin{theorem}\label{Th:main}
Specify two ideals $I=\langle f_1,\ldots,f_v \rangle $ and $J=\langle g_1,\ldots,g_u \rangle$ in $A$ with $I$ zero-dimensional. Let $>$ be a semigroup order on $\textup{T}^{\{x_1,\ldots,x_{n+u}\}}$ such that $x_1 >1, \ldots, x_n >1$ and $x_{n+1}<1, \ldots, x_{n+u}<1$. Consider the ideal $I'=\langle f_1,\ldots,f_v,g_1-x_{n+1},\ldots,g_u-x_{n+u} \rangle \subset \mathbb{C}[x_1,\ldots,x_{n+u}]$. Then, the complex zeros of $I'^{ec}=(\textup{Loc}_{>}(I'))^c$ correspond to the complex zeros of $I$ in variety $\textbf{V}(J)$ and corresponding zeros have the same multiplicities and first $n$ coordinates.
\end{theorem}

\begin{pf*}{Proof}
The complex zeros of $I$ in variety $\textbf{V}(J)$ correspond to the complex zeros of $I'$ with $x_{n+1}=0,\ldots,x_{n+u}=0$. By Lemma \ref{CorrespondingZeros}, corresponding zeros of $I$ and $I'$ have the same first $n$ coordinates and the same multiplicities. By Theorem \ref{decomposition}, we can obtain that $I'^{ec}=(\textup{Loc}_{>}(I'))^c$ has a minimal primary decomposition $\cap _{i=1}^m Q_i$ where $m$ is the number of distinct complex zeros of $I'$ with $x_{n+1}=0,\ldots,x_{n+u}=0$, $Q_i$ are primary ideals in $\mathbb{C}[x_1,\ldots,x_{n+u}]$ such that $\sqrt{Q_i}=P_i=\langle x_1-a_{i1},\ldots,x_n-a_{in},x_{n+1},\ldots,x_{n+u} \rangle$ and $(a_{i1},\ldots,a_{in},0,\ldots,0) \in \mathbb{C}^{n+u}$ are all complex zeros of $I'^{ec}$. By Corollary \ref{CorollaryMultiplicity}, for each $1 \leq i \leq m$, the multiplicity of $(a_{i1},\ldots,a_{in},0,\ldots,0) \in \mathbb{C}^{n+u}$ w.r.t. $I'^{ec}$ is the same as the multiplicity of the same point w.r.t. $I'$. Therefore, the zeros of $I'^{ec}=(\textup{Loc}_{>}(I'))^c$ correspond to the zeros of $I$ in variety $\textbf{V}(J)$ such that corresponding zeros have the same multiplicities and first $n$ coordinates.
\end{pf*}

\section{Computing Partial Solutions}\label{Sec:Computing}

In this section, we study how to use the standard basis method to count or compute the complex zeros of $I$ that locate in $\textbf{V}(J)$ with the multiplicities w.r.t. $I$.

\begin{theorem} \label{Th:Counting}
If $I \subset A$ is a zero-dimensional ideal and $>$ is a semigroup order on $\mathbb{Z}^n_{\geq 0}$, then $ A/I^{ec} \cong \textup{Loc}_{>}(A)/I\textup{Loc}_{>}(A)$.
\end{theorem}

\begin{pf}
$A/I^{ec} \cong \oplus_{i=1}^m A/Q_i$ and $\textup{Loc}_{>}(A)/I\textup{Loc}_{>}(A) \cong \oplus_{i=1}^m S^{-1}A/S^{-1}Q_i$ hold according to Theorem \ref{decomposition}. We only need to show $A/Q_i \cong S^{-1}A/S^{-1}Q_i$ for every $i=1,\ldots,m$. Let $\varphi_i: A \rightarrow S^{-1}A/S^{-1}Q_i$, $a \mapsto [a/1]$. This is a ring homomorphism. If we can prove $\ker(\varphi_i) = Q_i$ and $\varphi_i$ is onto, then we are done.

It is easy to see that $Q_i \subset \ker(\varphi_i)$. Note that $Q_i$ is primary and $S \cap \sqrt{Q_i}= \emptyset$. Then $(S^{-1}Q_i)^c=Q_i$ by Theorem \ref{Atiyah}. Let $b \in \ker(\varphi_i)$. Then $b/1 \in S^{-1}Q_i$ and $b \in (S^{-1}Q_i)^c$. We have $\ker(\varphi_i) \subset Q_i$. Thus, $\ker(\varphi_i)=Q_i$.

 Take an arbitrary element $[h/s]$ in $S^{-1}A/S^{-1}Q_i$ where $h \in A$ and $s \in S$. Let $P_i=\sqrt{Q_i}$ be the maximal ideal corresponding to a point $p_i$ in $\mathbb{C}^n$. By Theorem \ref{Th:index}, there exists a positive integer $k$ such that $P_i^k \subset Q_i$. Since $S \cap P_i =\emptyset$, we have $s(p_i)=c_0 \neq 0$. Hence, $1-s/c_0 \in P_i$ and $(1-s/c_0)^k \in P_i^k \subset Q_i \subset S^{-1}Q_i$. Take $d=(1+(1-s/c_0)+(1-s/c_0)^2+\cdots+(1-s/c_0)^{k-1}))/c_0 \in A$. Then $sd=c_0(1-(1-s/c_0))*(1+(1-s/c_0)+(1-s/c_0)^2+\cdots+(1-s/c_0)^{k-1}))/c_0=1-(1-s/c_0)^k \equiv 1 \mod S^{-1}Q_i$. Then $dh$ is a preimage of $[h/s]$, i.e., $\varphi_i$ is onto.
\end{pf}

\begin{corollary}\label{Cor:Counting}
If $I \subset A$ is a zero-dimensional ideal and $>$ is a semigroup order on $\mathbb{Z}^n_{\geq 0}$, then $ \dim A/I^{ec} = \dim \textup{Loc}_{>}(A)/I\textup{Loc}_{>}(A) < \infty$.
\end{corollary}

\begin{pf*}{Proof}
By Theorem \ref{decomposition}, $\dim A/I^{ec} = \sum _{i=1}^m \dim A/Q_i \leq \dim A/I < \infty$. Then by Theorem \ref{Th:Counting}, $\dim \textup{Loc}_{>}(A)/I\textup{Loc}_{>}(A)=\dim A/I^{ec}<\infty$.
\end{pf*}

Corollary \ref{Cor:Counting} transforms the problem of counting complex zeros of $I^{ec}$ with multiplicities to the problem of computing the $\mathbb{C}$-vector space dimension of $\textup{Loc}_{>}(A)/I\textup{Loc}_{>}(A)$. The next theorem shows that this dimension can be worked out by computing a standard basis of $I\textup{Loc}_{>}(A)$ in $\textup{Loc}_{>}(A)$ w.r.t. $>$. The number of the standard monomials is just the $\mathbb{C}$-vector space dimension of $\textup{Loc}_{>}(A)/I\textup{Loc}_{>}(A)$.
This theorem is an easy generalization of the one in \cite{Cox05}, where the order $>$ is a local order.
\begin{theorem}\label{Th:LT}
Let $>$ be a semigroup order on $\mathbb{Z}^n_{\geq 0}$ and $R$ be the ring $\textup{Loc}_{>}(A)$. If $I \subset R$ is an ideal, then the following are equivalent.
\begin{enumerate}
  \item $\dim R/I$ is finite.
  \item $\dim R/ \langle \textsc{lt}(I) \rangle$ is finite.
  \item There are only finitely many standard monomials.
\end{enumerate}
Furthermore, when any of these conditions is satisfied, we have $$\dim R/I = \dim R/ \langle \textsc{lt}(I) \rangle = \textup{number of standard monomials}$$ and every $f \in R$ can be written uniquely as a sum $$f=g+r,$$ where $g \in I$ and $r$ is a linear combination of standard monomials. In addition, this decomposition can be computed algorithmically in $R$.

\end{theorem}

\begin{pf}
The proof is almost the same as the proof of its local version in \cite{Cox05}. We only need to change a local order to a semigroup order and a local ring to $\textup{Loc}_{>}(A)$.
\end{pf}

In the proof of Theorem \ref{Th:LT}, a standard basis $G$ of $I$ w.r.t. $>$ should be computed first to reduce an element $f \in R$ to $r$ by the Mora normal form algorithm. Denote $r$ by $\textup{redNF}(f,G,>)$. Theorem \ref{Th:LT} and Corollary \ref{Cor:Counting} tell us that for a zero-dimensional ideal $I \subset A$ and an arbitrary semigroup order $>$ on $\mathbb{Z}^n_{\geq 0}$, there exists a unique reduced standard basis $G$ of $I^e=I\textup{Loc}_>(A)$ w.r.t. $>$. And for every $f \in \textup{Loc}_>(A)$ the reduced normal form $\textup{redNF}(f,G,>)$ is well defined.

By Theorem \ref{Th:main}, Corollary \ref{Cor:Counting} and Theorem \ref{Th:LT}, for a zero-dimensional polynomial ideal, we can count the number of its complex zeros that locate in another variety with multiplicities by computing a standard basis of a localization of a new polynomial ideal w.r.t. a semigroup ordering.

In the next theorem, we introduce a way to work out such partial solutions.

\begin{theorem} \label{Th:BorderBases}
If $I \subset A$ is a zero-dimensional ideal, $>$ is a semigroup order on $\mathbb{Z}^n_{\geq 0}$ and $G$ is a standard basis of $I\textup{Loc}_{>}(A)$ w.r.t. $>$, then there are only finitely many standard monomials. Moreover, these standard monomials form an order ideal $W$ and a $\mathbb{C}$-vector space basis of $A/I^{ec}$. Consequently, the set $H=\{ t-\textup{redNF}(t,G,>) : t \in \partial W \}$ is a $W$-border basis of the ideal $I^{ec}=(I\textup{Loc}_{>}(A))^c$ in $A$.
\end{theorem}

\begin{pf}
Since $I \subset A$ is zero-dimensional, we have $\dim \textup{Loc}_{>}(A)/I\textup{Loc}_{>}(A) <\infty$ by Corollary \ref{Cor:Counting}. Then, by Theorem \ref{Th:LT}, the set $W$ of standard monomials is finite. For every $t \in W$, the divisors of $t$ in $\textup{T}^{\{x_1,\ldots,x_n\}}$ is also in $W$ by Definition \ref{Def:StandardMonomial}. Thus, $W$ is an order ideal by Definition \ref{Def:OrderIdeal} and $H=\{ t-\textup{redNF}(t,G,>) : t \in \partial W \}$ is a $W$-border prebasis by Definition \ref{Def:Border} and Definition \ref{Def:Prebasis}. Note that $\dim A/ I^{ec} < \infty$ by Corollary \ref{Cor:Counting}. It implies $I^{ec}$ is zero-dimensional. By Theorem \ref{Th:LT}, $H \subset I\textup{Loc}_{>}(A) \cap A = I^{ec}$. Suppose there is a nontrivial linear combination $f$ of the monomials in $W$ with coefficients in $\mathbb{C}$ such that $f \in I^{ec} \subset I\textup{Loc}_{>}(A)$. Then the residue classes of these monomials in $\textup{Loc}_{>}(A)/I\textup{Loc}_{>}(A)$ are linearly dependent, a contradiction with the conclusion of Theorem \ref{Th:LT}. Hence, the residue classes of the monomials in $W$ in $A/I^{ec}$ are linearly independent. On the other hand, by Corollary \ref{Cor:Counting} and Theorem \ref{Th:LT}, $\dim A/I^{ec} =\dim \textup{Loc}_{>}(A)/I\textup{Loc}_{>}(A) =\#W$. As a result, we know that the residue classes of the monomials in $W$ form a $\mathbb{C}$-vector space basis of $A/I^{ec}$. Thus, by Definition \ref{Def:BorderBasis}, the set $H$ is a $W$-border basis of the ideal $I^{ec}$ in $A$.
\end{pf}

 According to Proposition \ref{Pro:BorderGenerators}, the border basis $H$ in Theorem \ref{Th:BorderBases} is a set of generators of $I^{ec}$. By using this border basis, we can construct multiplication matrices (cf. Definition \ref{Def:MultiplicationMatrix}) and a Chow form (cf. Defintion \ref{Def:ChowForm}). Then applying Theorem \ref{Th:ChowForm}, we can compute all the complex zeros of $I^{ec}$ in $\mathbb{C}^n$ with their multiplicities. Other methods can also be used to work out the zeros of $I^{ec}$ with multiplicities, for example, the rational univariate representation (RUR) method (cf. \cite{Rouillier99}).

\begin{example} \label{Example}
Milnor number is an important invariant in singularity theory. According to \cite{Greuel07}, for a holomorphic function $f$, the critical points are the points in $\mathbb{C}^n$ that vanish the ideal $I=\langle \partial f/\partial x_1,\ldots,\partial f/\partial x_n \rangle \subset A$, and the singular points are the critical points on the hypersurface $\textbf{V}(f)$. If $f$ is a polynomial in $A$, then the Milnor number of a critical point $p$ is just equal to the multiplicity of $p$ w.r.t. $I$ (note that not $\langle I,f \rangle$ for the singular points). We want to calculate the sum of Milnor numbers of the singular points on $\textbf{V}(f)$ and work out all the singular points with their Milnor numbers.

Take $f=(x^2+y^2-2y)(y-2x^2)=-2x^4-2x^2y^2+5x^2y+y^3-2y^2 \in A=\mathbb{C}[x,y]$ for example. We do the computation in a computer algebra software \textsc{Singular} (cf. \cite{DGPS} and \cite{Greuel07}). The codes and outputs are listed below.

\begin{tabular}{l}
\verb"ring B=0,(x,y,u),(a(1,1,-1),dp);" \\
\verb"poly f=-2x4-2x2y2+5x2y+y3-2y2;"\\
\verb"ideal I=jacob(f);"\\
\verb"ideal K=I,u-f;"\\
\verb"ideal G=std(K);"\\
\verb"G;"\\
\verb"G[1]=18xy2-27xy+64xu"\\
\verb"G[2]=45x2+9y2-32x2u-36y-27u"\\
\verb"G[3]=u"\\
\verb"G[4]=4y3+25x2-y2-20y-16u"
\end{tabular}

In the codes, $B$ is a ring $k[x,y,u]$ where $k$ is a field with characteristic $0$. For our purpose, we can view $B$ as $\mathbb{C}[x,y,u]$. The set $G$ is a standard basis of $\textup{Loc}_>(\langle I,u-f \rangle$) in $\textup{Loc}_>(B)$ w.r.t. a semigroup order $>$ such that $x>1$, $y>1$ and $u<1$. The \verb"dp" in the first line of the codes is an ordering chosen randomly to break ties. Note that the polynomials in $G$ are not reduced. All the terms in $G[1]$, $G[2]$ and $G[4]$ that contain factor $u$ can be cancelled by $G[3]$. $G[4]$ can be further reduced to $2y^3-3y^2$.

\begin{tabular}{l}
\verb"vdim(G);"\\
\verb"5"
\end{tabular}

The command \verb"vdim" computes the $\mathbb{C}$-vector space dimension of $\textup{Loc}_>(B)/\langle G \rangle$ for a standard basis $G$ w.r.t. $>$. Here, the output means the sum of Milnor numbers of singular points on $\textbf{V}(f)$ is $5$ (the sum of Milnor numbers of all the critical points is $8$). This result is based on Theorem \ref{Th:main}, Corollary \ref{Cor:Counting} and Theorem \ref{Th:LT}.

From the output of $G$ we can easily see that $1,y,x,y^2,xy$ form an order ideal and a $\mathbb{C}$-vector space basis of $B/\langle I,u-f \rangle^{ec}$. Compute the normal form of $x^2y$ (note that here and the previous reductions should use the Mora normal form algorithm, because we are working in $\textup{Loc}_>(B)$). Then, by Theorem \ref{Th:BorderBases}, we get a border basis $H=\{ 2xy^2-3xy,5x^2+y^2-4y,u,2y^3-3y^2, 2x^2y-y^2 \}$ of $\langle I,u-f \rangle^{ec}$ in $B$. Construct the multiplication matrices of $B/\langle I,u-f \rangle^{ec}$ w.r.t. the $\mathbb{C}$-vector space basis $\{1,y,x,y^2,xy\}$ from $H$:
$$M_x=\left( \begin{array}{ccccc}
0\ \ & 0\ \ & 1 & 0 & 0\\
0 & 0 & 0 & 0 & 1\\
0 &4/5& 0&-1/5& 0\\
0 & 0 & 0 & 0 &3/2\\
0 & 0 & 0 &1/2&0
\end{array}\right ),
M_y=\left( \begin{array}{ccccc}
0\ \ & 1\ \ & 0\ \ & 0 \ & 0\\
0 & 0 & 0 & 1 & 0\\
0 & 0 & 0 & 0 & 1\\
0 & 0 & 0 & 3/2 & 0\\
0 & 0 & 0 & 0 & 3/2
\end{array} \right ),
M_u=0_{5 \times 5}.$$\\
By Definition \ref{Def:ChowForm}, the Chow form of $B/\langle I,u-f \rangle^{ec}$ in $v=(v_0,v_1,v_2,v_3)$ is
\begin{eqnarray*}
\quad  \quad \quad \  C_{\langle I,u-f \rangle^{ec}}(v) & = & \det(v_0+v_1M_x+v_2M_y+v_3M_u) \\
  & = &v_0^3(v_0-\frac{\sqrt{3}}{2}v_1+\frac{3}{2}v_2)(v_0+\frac{\sqrt{3}}{2}v_1+\frac{3}{2}v_2).
\end{eqnarray*}
Therefore, there are three distinct singular points $(0,0)$, $(-\sqrt{3}/2,3/2)$ and $(\sqrt{3}/2,3/2)$ on the hypersurface $\textbf{V}(f)$ with respective Milnor numbers $3$, $1$ and $1$ by Theorem \ref{Th:ChowForm} and Theorem \ref{Th:main}. Note that $u$ does not appear in the Chow form since for each complex zero of $\langle I,u-f \rangle^{ec}$, its $u$-coordinate is $0$ by Theorem \ref{decomposition}. Thus, in fact, we do not need to compute $M_u$ for computing the Chow form.
\end{example}

\section{Conclusion}\label{Sec:Conclusion}

In this paper, we proposed a natural geometric explanation of the effect of a semigroup order $>$ on localizing polynomial rings $\mathbb{C}[x_1,\ldots,x_n]$, i.e., for an arbitrary zero-dimensional ideal $I$ in $\mathbb{C}[x_1,\ldots,x_n]$ and every $1 \leq i \leq n$, if $x_i <1$ then the complex zeros of $I$ with the $i$-th coordinates nonzero are discarded when we construct $I^{ec}$. Then, based on this view, we proved that for a zero-dimensional ideal, the standard basis method and the border basis method can be used to compute its complex zeros that locate in another variety. As an application, we computed singular points with their Milnor numbers on a polynomial hypersurface $\textbf{V}(f)$.

\bibliographystyle{elsart-harv}
\bibliography{JSC}

\begin{thebibliography}{14}
\expandafter\ifx\csname natexlab\endcsname\relax\def\natexlab#1{#1}\fi
\expandafter\ifx\csname url\endcsname\relax
  \def\url#1{\texttt{#1}}\fi
\expandafter\ifx\csname urlprefix\endcsname\relax\def\urlprefix{URL }\fi

\bibitem[{Atiyah and MacDonald(1969)}]{Atiyah1969}
Atiyah, M.~F., MacDonald, I.~G., 1969. Introduction to Commutative Algebra.
  Addison-Wesley.

\bibitem[{Buchberger(1965)}]{Buchberger65}
Buchberger, B., 1965. Ein Algorithmus zum Auffinden der Basiselemente des
  Restklassenringes nach einem nulldimensionalen Polynomideal. Ph.D. Thesis,
  Mathematical Institute, University of Innsbruck, Austria.

\bibitem[{Cox et~al.(2005)Cox, Little, and O'Shea}]{Cox05}
Cox, D., Little, J., O'Shea, D., 2005. Using Algebraic Geometry (Second
  Edition). Springer, USA.

\bibitem[{Decker et~al.(2012)Decker, Greuel, Pfister, and Sch\"onemann}]{DGPS}
Decker, W., Greuel, G.-M., Pfister, G., Sch\"onemann, H., 2012. {\sc Singular}
  {3-1-6} --- {A} computer algebra system for polynomial computations.
  \url{http://www.singular.uni-kl.de}.

\bibitem[{Faug\`ere and Liang(2011)}]{FL11Extended}
Faug\`ere, J.-C., Liang, Y., June 2011. Pivoting in extended rings for
  computing approximate {Gr\"obner} bases. Mathematics in Computer Science
  5~(2), 179--194.

\bibitem[{Gr\"abe(1994)}]{Graebe94}
Gr\"abe, H.-G., December 1994. The tangent cone algorithm and homogenization.
  Journal of Pure and Applied Algebra 97~(3), 303--312.

\bibitem[{Greuel and Pfister(1996)}]{Greuel96}
Greuel, G.-M., Pfister, G., February 1996. Advances and improvements in the
  theory of standard bases and syzygies. Archiv der Mathematik 66~(2),
  163--176.

\bibitem[{Greuel and Pfister(2008)}]{Greuel07}
Greuel, G.-M., Pfister, G., 2008. A \textsc{Singular} Introduction to
  Commutative Algebra (Second, Extended Edition). Springer, New York.

\bibitem[{Kehrein et~al.(2005)Kehrein, Kreuzer, and Robbiano}]{KKR}
Kehrein, A., Kreuzer, M., Robbiano, L., 2005. An algebraist's view on border
  bases. In: Dickenstein, A., Emiris, I.~Z. (Eds.), Solving Polynomial
  Equations: Foundations, Algorithms, and Applications. Algorithms and
  Computation in Mathematics. Springer, Ch.~4, pp. 160--202.

\bibitem[{Mora(1982)}]{Mora82}
Mora, F., 1982. An algorithm to compute the equations of tangent cones. In:
  {Proceedings of EUROSAM82}. Lecture Notes in Computer Science 144, Springer,
  Berlin, pp. 158--165.

\bibitem[{Mora and Rossi(1995)}]{Mora95}
Mora, T., Rossi, M.~E., 1995. An algorithm for the {Hilbert-Samuel} function of
  a primary ideal. Communications in Algebra 23~(5), 1899--1911.

\bibitem[{Mourrain(2007)}]{mou07}
Mourrain, B., 2007. Pythagore's dilemma, symbolic-numeric computation, and the
  border basis method. In: Wang, D., Zhi, L. (Eds.), Proceedings of {SNC 2007}.
  Birkh\"auser, Xi'an, pp. 223--243.

\bibitem[{Robbiano(1985)}]{Robbiano1985}
Robbiano, L., 1985. Term orderings on the polynomial ring. In: {Proceedings of
  EUROCAL1985}. Lecture Notes in Computer Science 204, Springer, New York, pp.
  513--517.

\bibitem[{Rouillier(1999)}]{Rouillier99}
Rouillier, F., May 1999. Solving zero-dimensional systems through the rational
  univariate representation. Journal of Applicable Algebra in Engineering,
  Communication and Computing 9~(5), 433--461.

\end{thebibliography}

\end{document}